\documentclass[runningheads]{llncs}

\usepackage{tikz}
\usetikzlibrary{arrows,positioning,shapes.geometric}
\usepackage{verbatim}
\usepackage{pgfplots}
\usepackage{colortbl}
\usepackage{amssymb}
\usepackage{multirow}
\usepackage{mathtools}
\usepackage{amsmath}
\usepackage{comment}
\usepackage{enumitem}
\usepackage{textcomp}
\usepackage{subcaption}
\usepackage{lscape}
\usepackage{subcaption}

\usepackage{perpage} 
\MakePerPage{footnote} 

\usepackage[hyphens]{url}

\usepackage{graphicx}
\usepackage{hyperref}
\newcommand{\orcid}[1]{\href{https://orcid.org/#1}{\includegraphics[scale=0.7]{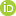}}}

\usepackage{soul}
\soulregister\ref7
\soulregister\cite7

\usepackage{longtable}
\begin{document}
\definecolor{tabhead}{RGB}{220,220,220}

\title{Machine Learning Based IoT Intrusion Detection System: An MQTT Case Study (MQTT-IoT-IDS2020 Dataset)}
\titlerunning{ML for MQTT IDS}
%
%
\author{Hanan~Hindy\inst{1}\orcidID{0000-0002-5195-8193} \and 
Ethan~Bayne\inst{1}\orcidID{0000-0003-1853-2921} \and
Miroslav~Bures\inst{2}\orcidID{0000-0002-2994-7826}
\and 
Robert~Atkinson\inst{3}\orcidID{0000-0002-6206-2229} 
\and  
Christos~Tachtatzis\inst{3}\orcidID{0000-0001-9150-6805} 
\and 
Xavier~Bellekens\inst{3}\orcidID{0000-0003-1849-5788}}
\authorrunning{H. Hindy et al.}
%
\institute{Division of Cyber Security, Abertay University, Dundee, Scotland, UK 
\email{\{1704847,e.bayne\}@abertay.ac.uk} \\ 
\and
Department of Computer Science, FEE, Czech Technical University in Prague, Czechia \\ 
\email{miroslav.bures@fel.cvut.cz} \\
\and 
EEE Department, University of Strathclyde, Glasgow, Scotland, UK
\email{\{robert.atkinson,christos.tachtatzis,xavier.bellekens\}@strath.ac.uk}}

\maketitle              
\begin{abstract}
The Internet of Things~(IoT) is one of the main research fields in the Cybersecurity domain. This is due to (a)~the increased dependency on automated device, 
and (b)~the inadequacy of general-purpose Intrusion Detection Systems~(IDS) to be deployed for special purpose networks usage. Numerous lightweight  protocols are being proposed for IoT devices communication usage.  One of the distinguishable IoT machine-to-machine communication protocols is Message Queuing Telemetry Transport~(MQTT) protocol. However, as per the authors best knowledge, there are no available IDS datasets that include MQTT benign or attack instances and thus, no IDS experimental results available. 

In this paper, the effectiveness of six Machine Learning~(ML) techniques to detect MQTT-based attacks is evaluated. Three abstraction levels of features are assessed, namely, packet-based, unidirectional flow, and bidirectional flow features. An MQTT simulated dataset is generated and used for the training and evaluation processes. The dataset is released with an open access licence to help the research community further analyse the accompanied challenges. 
The experimental results demonstrated the adequacy of the proposed ML models to suit MQTT-based networks IDS requirements. Moreover, the results emphasise on the importance of using flow-based features to discriminate MQTT-based attacks from benign traffic, while packet-based features are sufficient for traditional networking attacks.

\end{abstract}

\newcolumntype{C}[1]{>{\centering\arraybackslash}m{#1}}
\newcolumntype{L}[1]{>{\begin{math}}c<{\end{math}}}
\setlength{\belowcaptionskip}{-5pt}
\tabcolsep 2pt
\newdimen\NetTableWidthSix
 \NetTableWidthSix=\dimexpr
    \linewidth - 12\tabcolsep - 7\arrayrulewidth

\newdimen\NetTableWidthEight
 \NetTableWidthEight=\dimexpr
    \linewidth - 16\tabcolsep - 9\arrayrulewidth

\newdimen\NetTableWidthFour
 \NetTableWidthFour=\dimexpr
    \linewidth - 8\tabcolsep - 5\arrayrulewidth

\newdimen\NetTableWidthTen
 \NetTableWidthTen=\dimexpr
    \linewidth - 20\tabcolsep - 11\arrayrulewidth

\section{Introduction}
A large number of Internet of Things~(IoT) devices and networks have been utilised over the past years for different usage scenarios~\cite{hodo2016threat}. These use-cases include healthcare~\cite{10.1007/978-981-10-6875-1_66}, smart cities~\cite{arasteh2016iot}, supply~chain~\cite{abdel2018internet} and farming~\cite{ahmed2018internet}. With this extended use of IoT, new protocols are being deployed~\cite{nogues2019labelled}.
One of the new prominent protocols used for machine-to-machine communication is MQTT~\cite{stanford2013mqtt}. 

Harsha~\textit{et al.}~\cite{8554472} discuss the different protocols used in various IoT networks, which include MQTT. The authors analyse the security risks associated with using MQTT. The authors results show that there are 53396 publicly available and accessible MQTT devices~\cite{8554472}. Their work, alongside the work by Dinculean{\u{a}} and Cheng~\cite{dinculeanua2019vulnerabilities}, emphasises on the need for robust detection techniques for MQTT attacks to overcome the security vulnerabilities. 

As discussed in~\cite{8551386}, IoT Intrusion Detection Systems~(IDS) have different requirements due to the uniqueness of the usage scenarios involved. IoT IDSs are required to be flexible, extendable, and built using real or simulated traffic suited for the intended usage~\cite{9108270}. However, publicly available IoT datasets are limited, thus limiting IoT IDS development~\cite{9108270}.

In this manuscript, we aim at proposing and evaluating different Machine Learning~(ML) based MQTT IDS. The contributions of this paper are as follows:
\begin{itemize}
\item Generating a novel IoT 
-MQTT dataset and releasing it for public consumption.
\item Analysing a novel MQTT dataset which includes both benign and attack scenarios.
\item Evaluating the significance of using high-level (flow-based) features to build the IDS.
\item Assessing the proposed model using six different ML techniques. 
\item Examining the different needs of MQTT-based versus generic attacks detection, which emphasise the special setup and, thus the needs of MQTT~(IoT) networks.
\end{itemize}

The remainder of this paper is organised as follows; Section~\ref{sec:dataset} discusses the setup used for the dataset generation and provides an overview of the dataset and the extracted features. Section~\ref{sec:results} presents the results obtained by applying different ML techniques to detect attacks. Finally, the paper is concluded in Section~\ref{sec:conclusion}.
\section{Dataset}
\label{sec:dataset}
This section provides a description of the dataset gathered by the MQTT sensors simulation. The dataset is published\footnote{\url{https://ieee-dataport.org/open-access/mqtt-internet-things-intrusion-detection-dataset}} in~\cite{mqttdataset}. The dataset consists of five recorded scenarios; normal operation and four attack scenarios. The attacker performs four attack and each is recorded independently. 

\vspace{3mm}
\noindent
The attack types are: 
\begin{itemize}
    \item Aggressive scan~(Scan\_A)
    \item User Datagram Protocol~(UDP) scan~(Scan\_sU)
    \item Sparta SSH brute-force~(Sparta)
    \item MQTT brute-force attack~(MQTT\_BF)
\end{itemize}

\noindent
The data is acquired using tcpdump. The packets are collected by recording Ethernet traffic and then exporting to pcap files. The following tools were used as follows:
\begin{itemize}
    \item Virtual machines are used to simulate the network devices.
    \item Nmap is used for the scanning attacks.
    \item VLC is used to simulate the camera feed stream.
    \item MQTT-PWN~\cite{pwn} is used for the MQTT brute-force attack.
\end{itemize}

\begin{figure}[!hb]
	\centering
    \includegraphics[width=\textwidth]{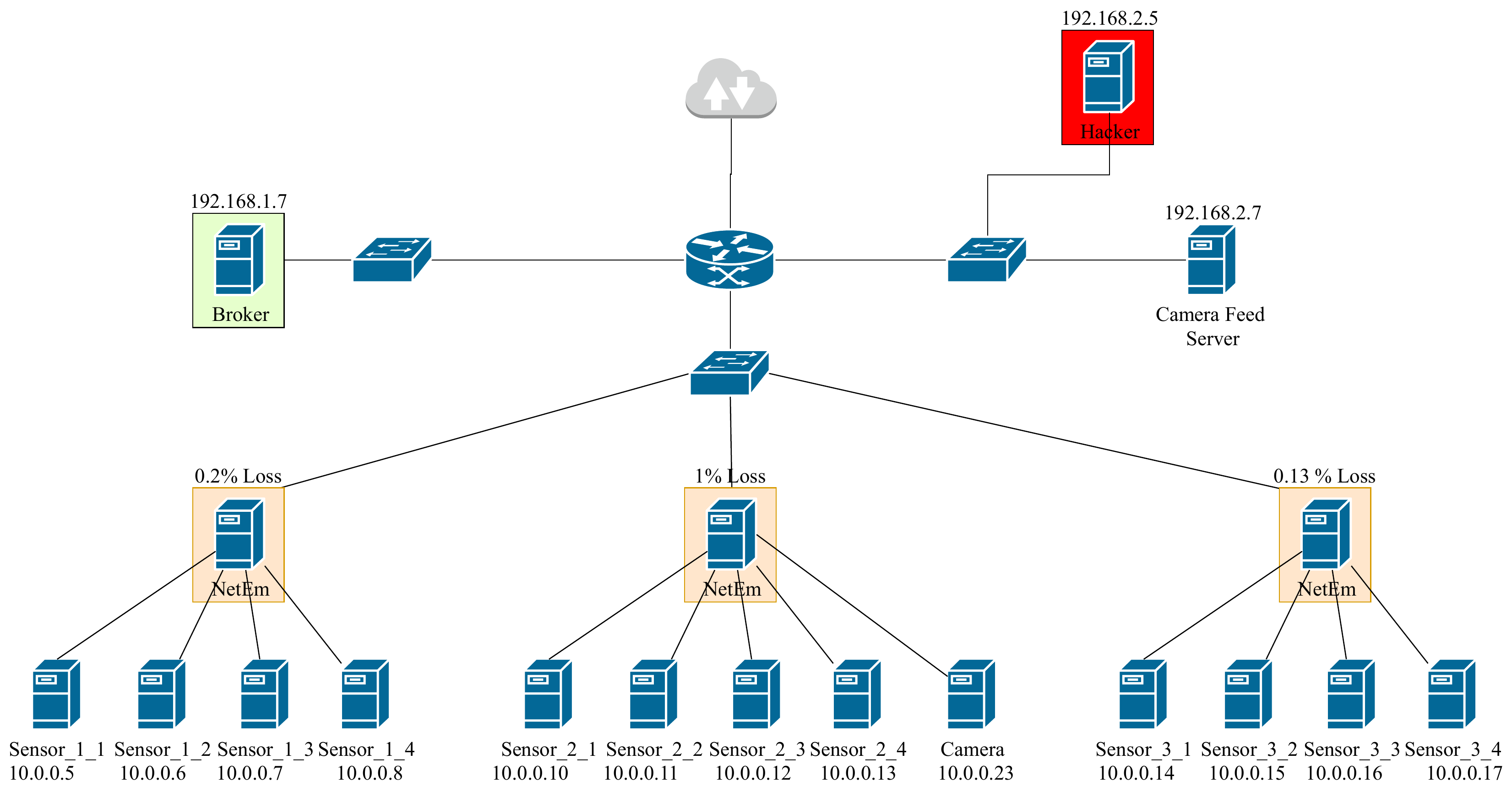}
    \caption{MQTT Network Architecture~\cite{mqttdataset}}
    \label{fig:networkArchitecture}
\end{figure}

Figure~\ref{fig:networkArchitecture} visualises the network components. The network consists of 12 MQTT sensors, a broker, a machine to simulate camera feed, and an attacker. During normal operation, all 12 sensors send randomised messages using the ``Publish'' MQTT command. The length of the messages is different between sensors to simulate different usage scenarios. The messages content is randomly generated. 
The camera feed is simulated using VLC media player which uses UDP stream. To further simulate a realistic scenario each of the network emulators drop packets with 0.2\%, 1\%, and 0.13\%. During the four attack scenarios recording, the background normal operation was left in action. The operating systems of the different devices are as follows; Tiny Core Linux for the sensors, Ubuntu for the camera \& camera feed server, and finally, Kali Linux for the hacker. 

The importance of this dataset is fourfold:
\begin{itemize}
    \item The dataset simulates a realistic MQTT IoT network in a normal operation scenario.
    \item The dataset includes both generic networking scanning attacks, as well as, MQTT brute-force attack.
    \item Researchers can use this dataset to build and evaluate IoT Intrusion Detection Systems.
    \item The dataset is the first to include MQTT scenarios and attacks data.
\end{itemize}

The dataset is provided in its raw capture format (.pcap files), as well as processed features~\cite{mqttdataset}. The features represent: (a)~packet-based features, (b)~Unidirectional-based features, and (c)~bidirectional-based features~\cite{Ring2017}. Each feature set is used exclusively, as discussed in Section~\ref{sec:results}.
The basic packet extracted features are listed in Table~\ref{tab:features}, fourth column. The feature list for unidirectional and bidirectional is listed in Table~\ref{tab:features}, columns five and six, respectively. It is important to note that for the bidirectional flows, some features (pointed as *) have two values---one for the forward flow and one for the backward flow. The two features are recorded and distinguished by a prefix ``fwd\_'' for forward and ``bwd\_'' for backward~\cite{mqttdataset}. Furthermore, the distribution of instances is listed in Table~\ref{tab:instances}. 

In order to avoid specific features influence, the following features are dropped. These features are source and destination IP addresses, protocol, and MQTT flags. The data is split into 75\% and 25\% for training and testing, respectively. 

\begin{longtable}[!h]{|C{0.27\NetTableWidthSix}|C{0.12\NetTableWidthSix}|C{0.36\NetTableWidthSix}|C{0.09\NetTableWidthSix}|C{0.08\NetTableWidthSix}|C{0.08\NetTableWidthSix}|}
 \caption{Features Description\label{tab:features}}\\
 \hline
\rowcolor{gray!30}
 Feature & Data Type & Description & Packet & Uni-flow & Bi-flow \\ \hline
 \endfirsthead

 \hline
 \multicolumn{6}{|c|}{Table \ref{tab:features} continued}\\
 \hline
\rowcolor{gray!30}
 Feature & Data Type & Description & Packet & Uni-flow & Bi-flow \\ \hline
 \hline
 \endhead

 ip\_src & Text & Source IP Address & \checkmark  & \checkmark &  \checkmark \\ \hline
ip\_dest & Text & Destination IP Address & \checkmark  & \checkmark &  \checkmark \\ \hline
protocol &	Text & Last layer protocol & \checkmark & & \\ \hline
ttl	& Integer & Time to live &	\checkmark & & \\ \hline 		
ip\_len	& Integer &	Packet Length & \checkmark & & \\ \hline
ip\_flag\_df & Binary &	Don't fragment IP flag &	\checkmark 	& & \\ \hline
ip\_flag\_mf & Binary &	More fragments IP flag &	\checkmark  & &	\\ \hline
ip\_flag\_rb & Binary &	Reserved IP flag &	\checkmark & & 	\\ \hline
prt\_src	& Integer &	Source Port &	\checkmark 	& \checkmark  &	\checkmark \\ \hline
prt\_dst	& Integer &	Destination Port &	\checkmark & 	\checkmark &	\checkmark \\ \hline
proto	& Integer &	Transport Layer protocol (TCP/UDP) &	& \checkmark &	\checkmark \\ \hline
tcp\_flag\_res & Binary	& Reserved TCP flag & 	\checkmark	& & 	\\ \hline
tcp\_flag\_ns	& Binary	& Nonce sum TCP flag	& \checkmark & &	\\ \hline 
tcp\_flag\_cwr	& Binary	& Congestion Window Reduced TCP flag& 	\checkmark	& 	& \\ \hline
tcp\_flag\_ecn	& Binary	& ECN Echo TCP flag	& \checkmark	& &	\\ \hline
tcp\_flag\_urg	& Binary	& Urgent TCP flag	& \checkmark		& &\\ \hline
tcp\_flag\_ack	& Binary	& Acknowledgement TCP flag	& \checkmark& 	&	\\ \hline
tcp\_flag\_push	& Binary	& Push TCP flag	& \checkmark		& & \\ \hline
tcp\_flag\_reset	& Binary	& Reset TCP flag	& \checkmark& 	&	\\ \hline
tcp\_flag\_syn	& Binary	& Synchronization TCP flag	& \checkmark& &	\\ \hline	
tcp\_flag\_fin	& Binary	& Finish TCP flag	& \checkmark& & 		\\ \hline
num\_pkts	& Integer & 	Number of Packets in the flow		& & \checkmark &	* \\ \hline
mean\_iat	& Decimal &  	Average inter arrival time		& & \checkmark &	* \\ \hline
std\_iat	& Decimal &   Standard deviation of inter arrival time		& & \checkmark &	* \\ \hline
min\_iat	& Decimal &   Minimum inter arrival time		& & \checkmark &	* \\ \hline
max\_iat	& Decimal &   Maximum inter arrival time		& & \checkmark &	* \\ \hline
num\_bytes	& Integer & 	Number of bytes		& & \checkmark &	* \\ \hline
num\_psh\_flags	& Integer &  Number of push flag		& & \checkmark &	* \\ \hline
num\_rst\_flags	& Integer & 	Number of reset flag		& & \checkmark &	* \\ \hline
num\_urg\_flags	& Integer & 	Number of urgent flag		& & \checkmark &	* \\ \hline
mean\_pkt\_len	& Decimal &   Average packet length		& & \checkmark &	* \\ \hline
std\_pkt\_len	& Decimal &  	Standard deviation packet length		& & \checkmark &	* \\ \hline
min\_pkt\_len	& Decimal &  	Minimum packet length		& & \checkmark &	* \\ \hline
max\_pkt\_len	& Decimal &  	Maximum packet length		& & \checkmark &	* \\ \hline
mqtt\_messagetype   	& Integer  &	MQTT message type	& \checkmark & & \\\hline		
mqtt\_messagelength
	& Binary &	MQTT message length	& \checkmark & & \\\hline		
mqtt\_flag\_uname	& Binary &	User Name MQTT Flag	& \checkmark & & \\\hline		
mqtt\_flag\_passwd	& Binary &	Password MQTT flag	& \checkmark & & \\\hline		
mqtt\_flag\_retain	& Binary &	Will retain MQTT flag	& \checkmark & & \\\hline		
mqtt\_flag\_qos	& Integer  &	Will QoS MQTT flag	& \checkmark & & \\\hline		
mqtt\_flag\_willflag	& Binary &	Will flag MQTT flag	& \checkmark & & \\\hline		
mqtt\_flag\_clean	& Binary &	Clean MQTT flag	& \checkmark & & \\\hline		
mqtt\_flag\_reserved	& Binary &	Reserved MQTT flag	& \checkmark & & \\\hline		
is\_attack	& Binary & 	1 if the instance represents an attack, 0 otherwise.& 	x &	x& 	x \\ \hline
\rowcolor{gray!30}
\multicolumn{6}{C{\textwidth}}{* represented as two features in the biflow features file (forward fwd and backward bwd)}
 \end{longtable}

\begin{table}[htb]
    \centering
    \caption{Dataset Instances Distribution}
    \begin{tabular}{|C{0.15\NetTableWidthEight}|C{0.11\NetTableWidthEight}|C{0.11\NetTableWidthEight}|C{0.11\NetTableWidthEight}|C{0.11\NetTableWidthEight}|C{0.11\NetTableWidthEight}|C{0.11\NetTableWidthEight}|C{0.11\NetTableWidthEight}|}
        \hline
        \rowcolor{gray!30}
         File Name &  pcap file size & \multicolumn{2}{C{0.235\NetTableWidthEight}|}{Number of Packets} & \multicolumn{2}{C{0.235\NetTableWidthEight}|}{Number of Uni-flow Instances} & \multicolumn{2}{C{0.235\NetTableWidthEight}|}{Number of Uni-flow Instances}\\ \cline{3-8}
         \rowcolor{gray!30}
          & & Benign & Attack &  Benign & Attack &  Benign & Attack \\ \hline
         \cellcolor{gray!30} normal & 192.5 MB	& 1056230 \newline (3.42\%)	& 0 &	171836  \newline (59.01\%)	& 0 & 	86008 \newline (54.78\%) &	0 \\ \hline
         \cellcolor{gray!30} scan\_A \newline (aggressive)	& 16.2 MB &	70768 &	40624 \newline (0.13\%) & 11560	& 39797 \newline (13.67\%)	&  5786 &	19907 \newline (12.68\%) \\\hline 
          \cellcolor{gray!30}scan\_sU \newline (UDP) &	41.3 MB	& 210819	& 	22436 \newline (0.07\%)	& 	34409	& 	22436 \newline (7.71\%)		& 17230	& 	22434 \newline (14.29\%) \\ \hline
            \cellcolor{gray!30} sparta	& 	3.4 GB	& 	947177		& 19728943 \newline (63.93\%)	& 	154175	& 	28232 \newline (9.7\%)	& 	77202	& 	14116 \newline (8.99\%) \\\hline

    \end{tabular}
    \label{tab:instances}
\end{table}

\section{Experiments and Results}
\label{sec:results}
This section discusses the conducted experiments. Note that the code is available on a GitHub repository \footnote{\url{https://github.com/AbertayMachineLearningGroup/MQTT\_ML}}.

Five-fold cross validation is used to evaluate each experiment. The metrics used for evaluation are as follows~\cite{9108270}: (a) Overall accuracy, as defined in equation~\ref{eq:accuracy}, such that True Positive~(TP) represents the attack instances correctly classified, True Negative~(TN) represents the benign instances correctly classified, Positive~(P) represents the number of attack instances and Negative~(N) represents the total number of benign instance.
\begin{equation}
\label{eq:accuracy}
Overall Accuracy = \frac{TP + TN}{P + N}
\end{equation}

For each class, Precision, Recall, and F1 Score are computed as shown in Equation~\ref{eq:precision}, Equation~\ref{eq:recall}, and Equation~\ref{eq:f1}, respectively~\cite{9108270}. False Positive~(FP) represents benign instances falsely classified as attack and False Negative~(FN) represents the attack instances falsely classified as benign. 

\begin{equation} 
\label{eq:precision}
Precision = \frac{TP}{TP + FP}
\end{equation}

\begin{equation} 
\label{eq:recall}
Recall = \frac{TP}{TP + FN}
\end{equation}

\begin{equation} 
\label{eq:f1}
F1 = \frac{2TP}{2TP + FP + FN}
\end{equation}

Finally, the weighted average for precision, recall, and F1 score is calculated to demonstrate the overall performance.

Six ML techniques are employed for the classification purpose. The ML techniques are:  Logistic Regression~(LR), Gaussian Na{\"i}ve Bayes~(NB) , k-Nearest Neighbours~(k-NN) , Support Vector Machine~(SVM) , Decision Trees~(DT)  and Random Forests~(RF) \cite{RefWorks:doc:5af4661ee4b0dfb887b46a75}~\cite{RefWorks:doc:5af46286e4b02abf496e090c}~\cite{RefWorks:doc:5af4651ee4b02dfcb38d5a2b} \cite{RefWorks:doc:5af4638ae4b0f7bd1fabb685}~\cite{RefWorks:doc:5af464f7e4b0cfc1f4ac242d}~\cite{RefWorks:doc:5af4642de4b04303c30e488f}~\cite{10.1007/978-3-030-12786-2_1}.

\begin{figure}
    \centering
    \includegraphics[width=0.9\linewidth]{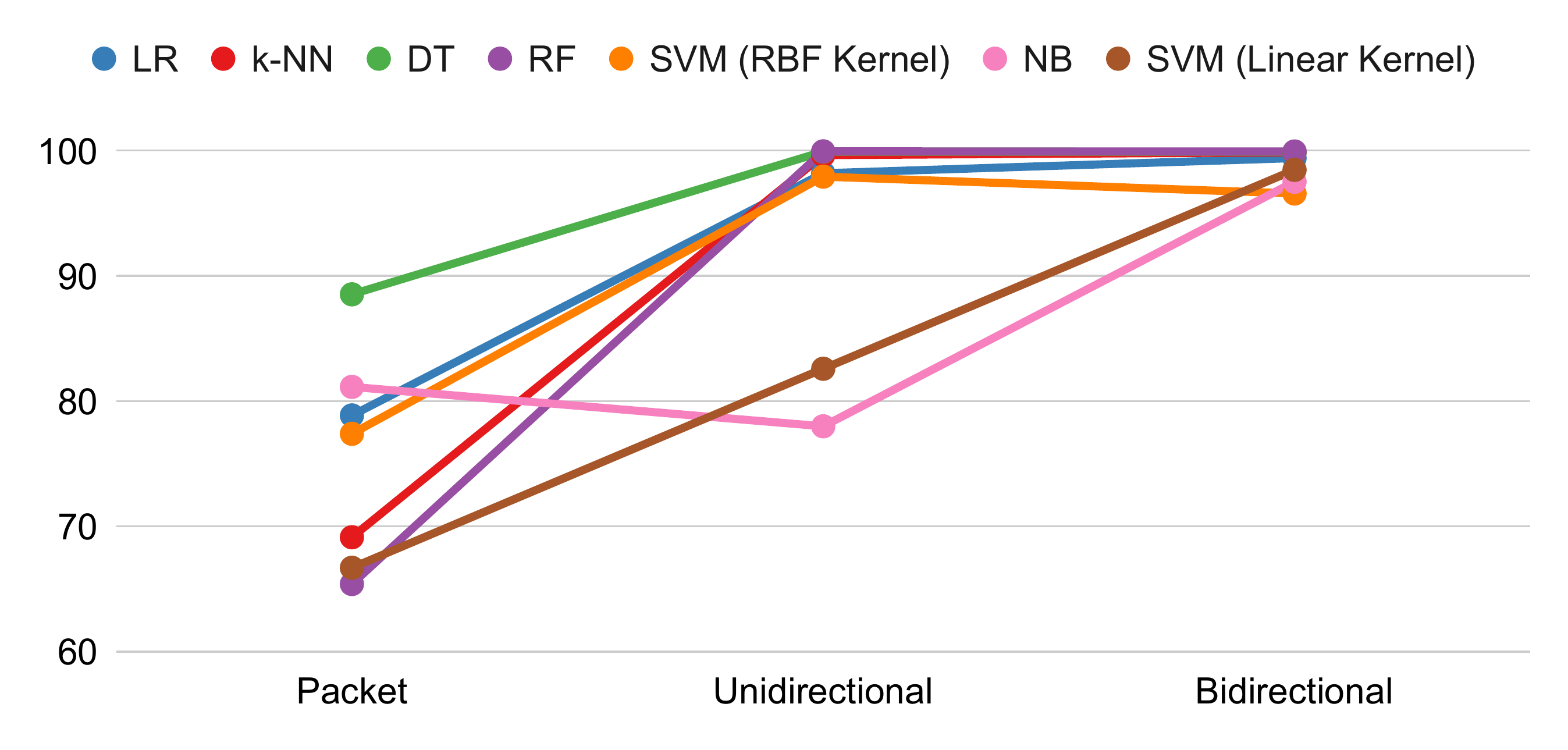}
    \caption{Overall detection accuracy trend using different ML techniques}
    \label{fig:overall-accuracy}
\end{figure}

Table~\ref{tab:overall-accuracy} details the overall accuracy of each of the ML techniques with packet, unidirectional and bidirectional features. It can be observed the performance rise accompanying flow-based features, both unidirectional and bidirectional. 
This rise could further be visualised in Figure~\ref{fig:overall-accuracy}.

\begin{table}[htb]
    \centering
    \caption{Overall detection accuracy}

    \begin{tabular}{| C{0.35\NetTableWidthFour} | C{0.2\NetTableWidthFour} | C{0.2\NetTableWidthFour} | C{0.2\NetTableWidthFour} |}
       \cline{2-4}
        \multicolumn{1}{c|}{} & \multicolumn{3}{c|}{\cellcolor{tabhead}\textbf{Features}} \\ \cline{2-4}
    	\multicolumn{1}{c|}{} & \cellcolor{tabhead}\textbf{Packet} & \cellcolor{tabhead}\textbf{Unidirectional} & \cellcolor{tabhead}\textbf{Bidirectional} \\ \hline
        \cellcolor{tabhead}\textbf{LR} & 78.87\% & 98.23\% & 99.44\% \\ \hline
        \cellcolor{tabhead}\textbf{k-NN} & 69.13\% & 99.68\% & 99.9\% \\ \hline
        \cellcolor{tabhead}\textbf{DT} & 88.55\% & 99.96\% & 99.95\% \\ \hline
        \cellcolor{tabhead}\textbf{RF} & 65.39\% & 99.98\% & 99.97\% \\ \hline
        \cellcolor{tabhead}\textbf{SVM (RBF Kernel)} & 77.4\% & 97.96\% & 96.61\% \\ \hline
        \cellcolor{tabhead}\textbf{NB} & 81.15\% & 78\% & 97.55\% \\ \hline
        \cellcolor{tabhead}\textbf{SVM (Linear Kernel)} & 66.69\% & 82.6\% & 98.5\% \\ \hline
    \end{tabular}
    \label{tab:overall-accuracy}
\end{table}

To further analyse the results, Table~\ref{tab:metrics1}, Table~\ref{tab:metrics2} and, Table~\ref{tab:metrics3} show the detailed precision, recall, and F1-score for each of the classifiers. 
Each classifier is represented as a sub-table. 
Similar to Table~\ref{tab:overall-accuracy}, it is observed that the flow-based features are strongly enhance the results. 

\begin{figure}[h]
\begin{subfigure}{.5\textwidth}
  \centering
  \includegraphics[width=\linewidth]{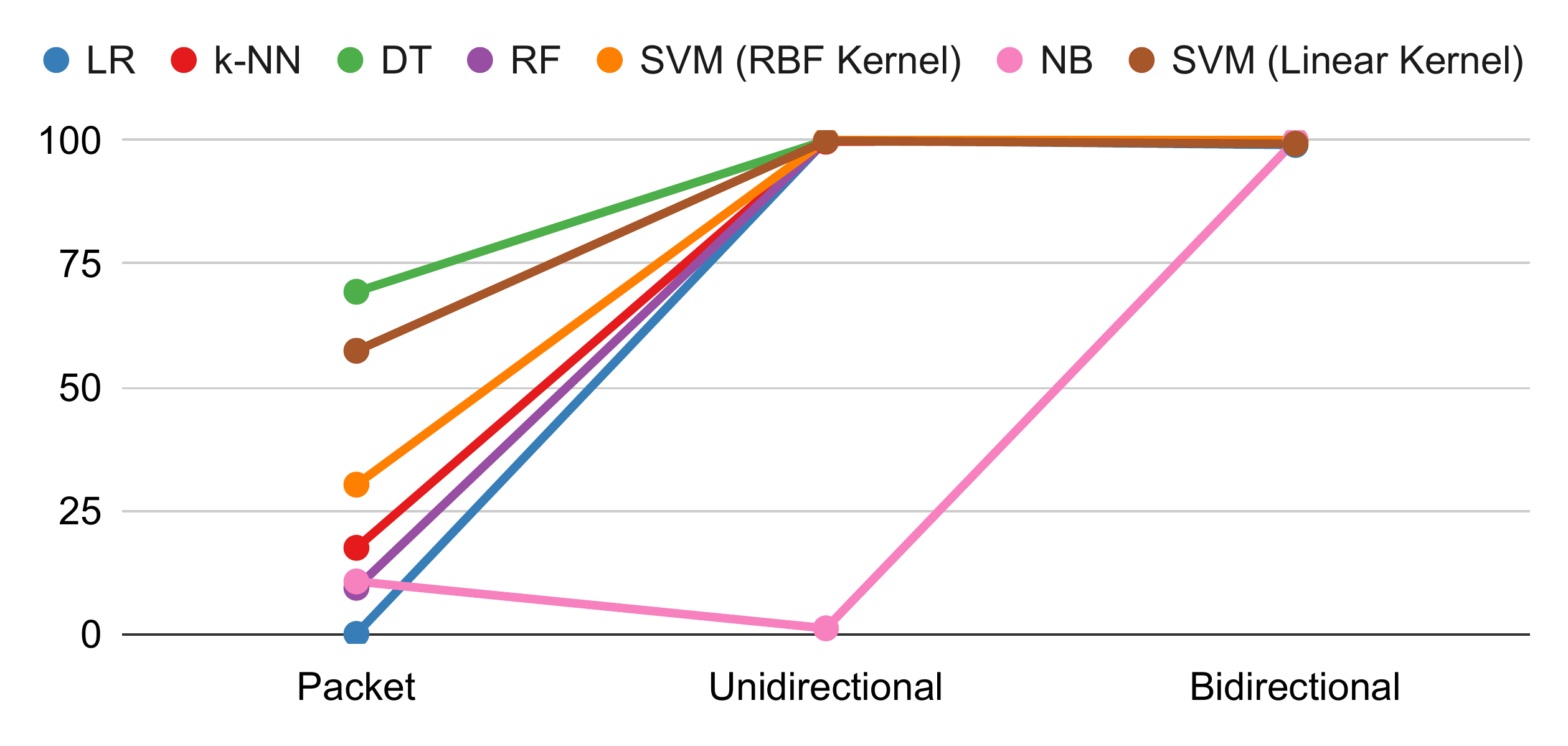}  
  \caption{Recall}
  \label{fig:recall-benign}
\end{subfigure}
\begin{subfigure}{.5\textwidth}
  \centering
  \includegraphics[width=\linewidth]{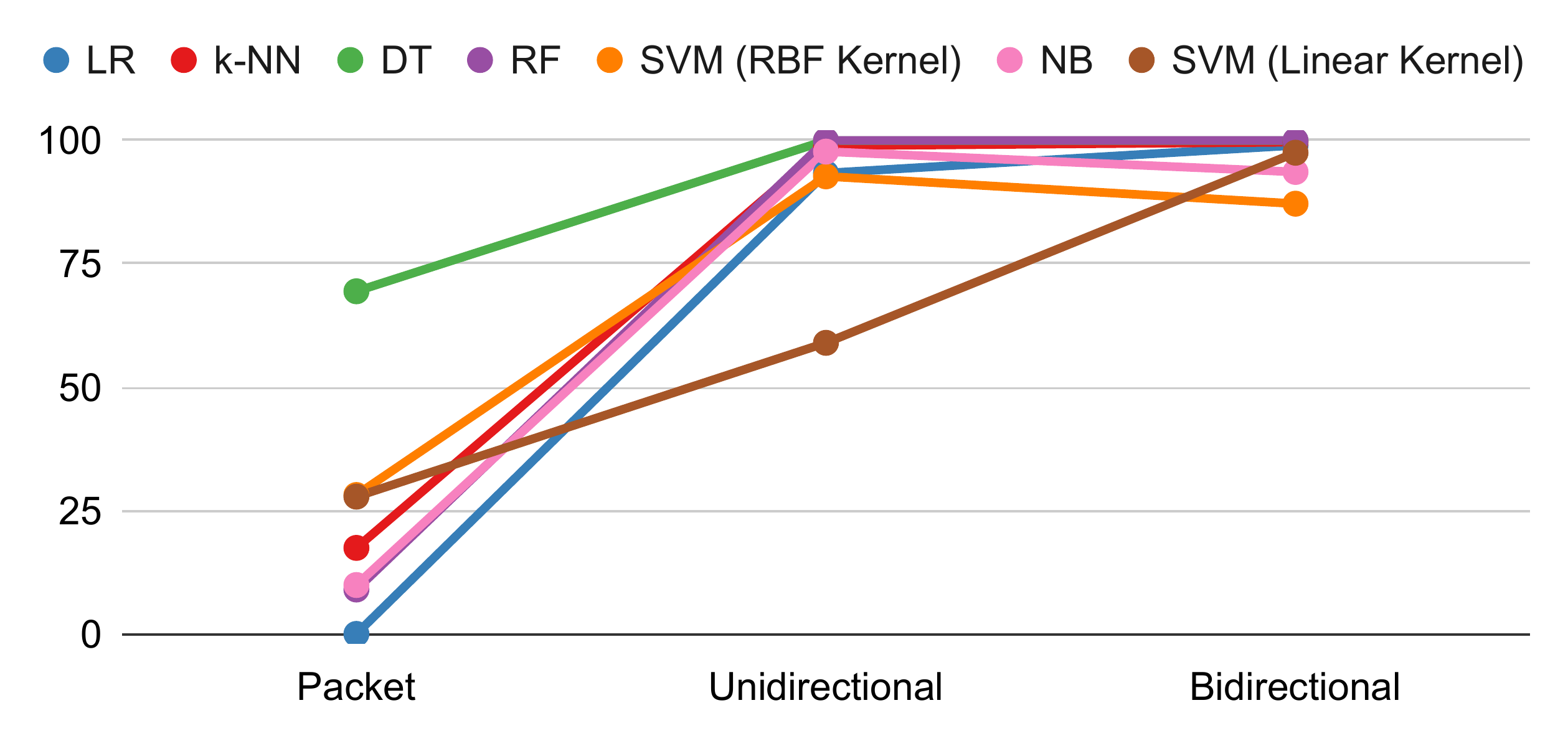}  
  \caption{Precision}
  \label{fig:precision-benign}
\end{subfigure}
\caption{Benign Class Trends}
\label{fig:benign-trend}
\end{figure}

\begin{figure}[ht]
\begin{subfigure}{.5\textwidth}
  \centering
  \includegraphics[width=\linewidth]{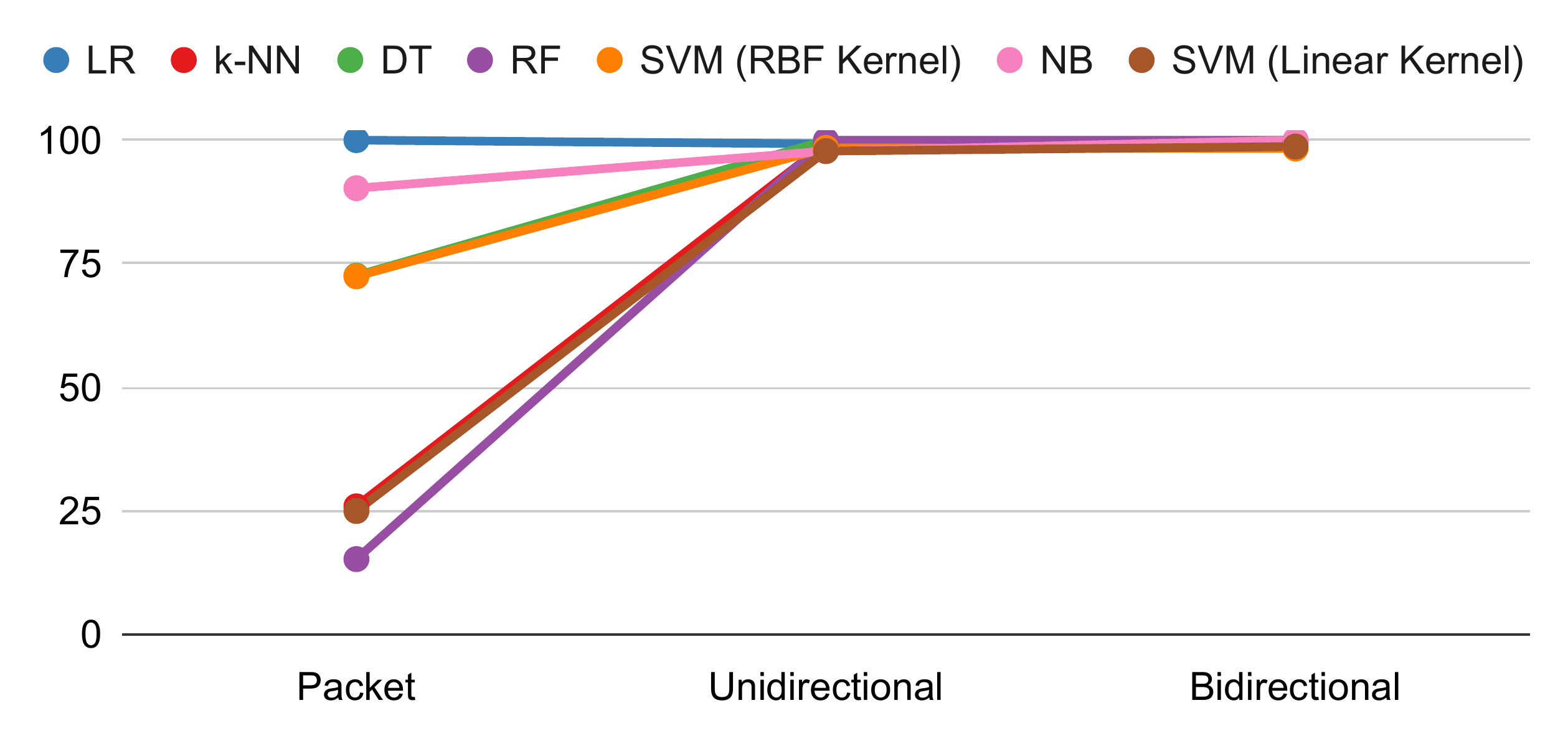}  
  \caption{Recall}
  \label{fig:recall-mqtt}
\end{subfigure}
\begin{subfigure}{.5\textwidth}
  \centering
  \includegraphics[width=\linewidth]{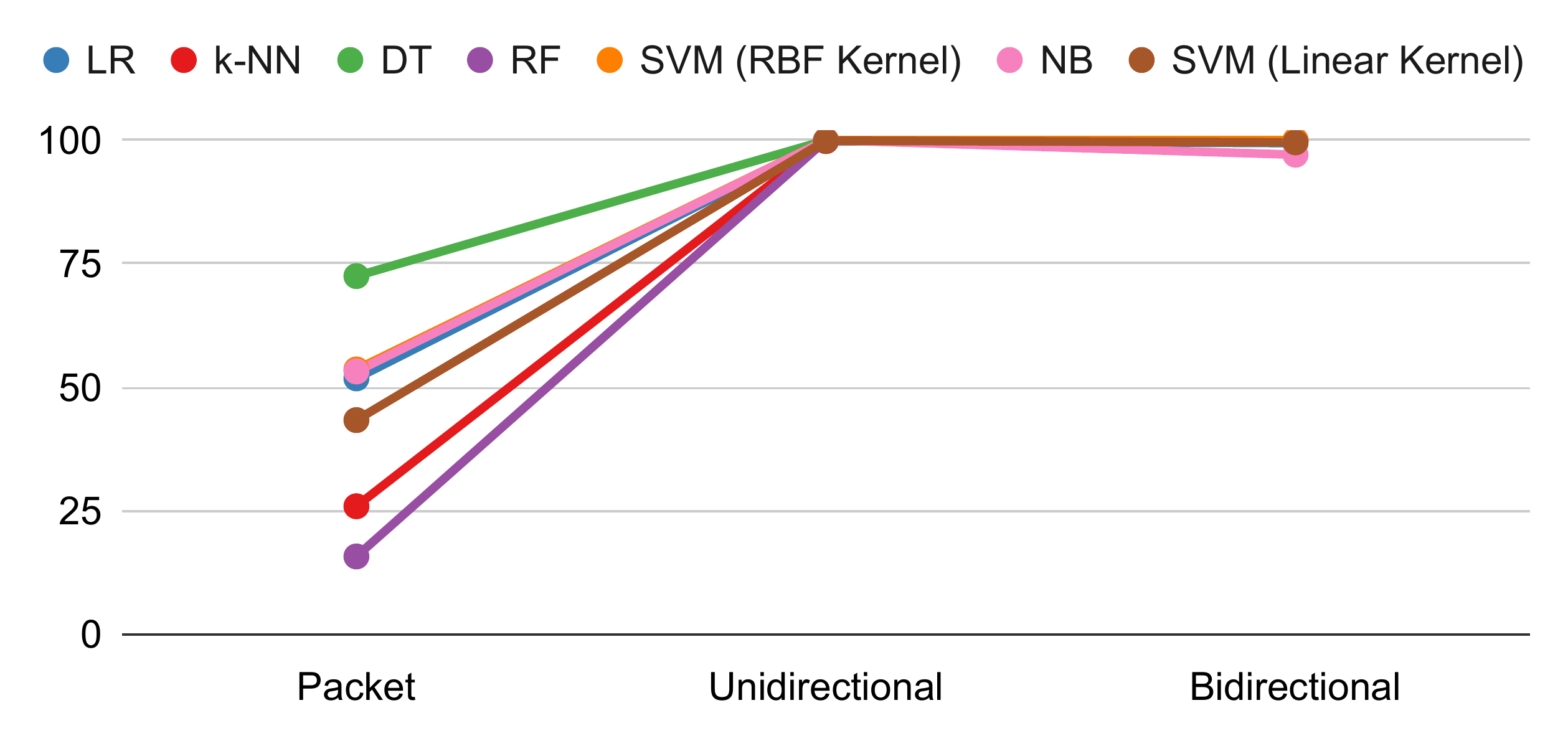}  
  \caption{Precision}
  \label{fig:precision-mqtt}
\end{subfigure}
\caption{MQTT\_BF Class Trends}
\label{fig:mqtt-trend}
\end{figure}

\begin{figure}[h!]
\begin{subfigure}{.5\textwidth}
  \centering
  \includegraphics[width=\linewidth]{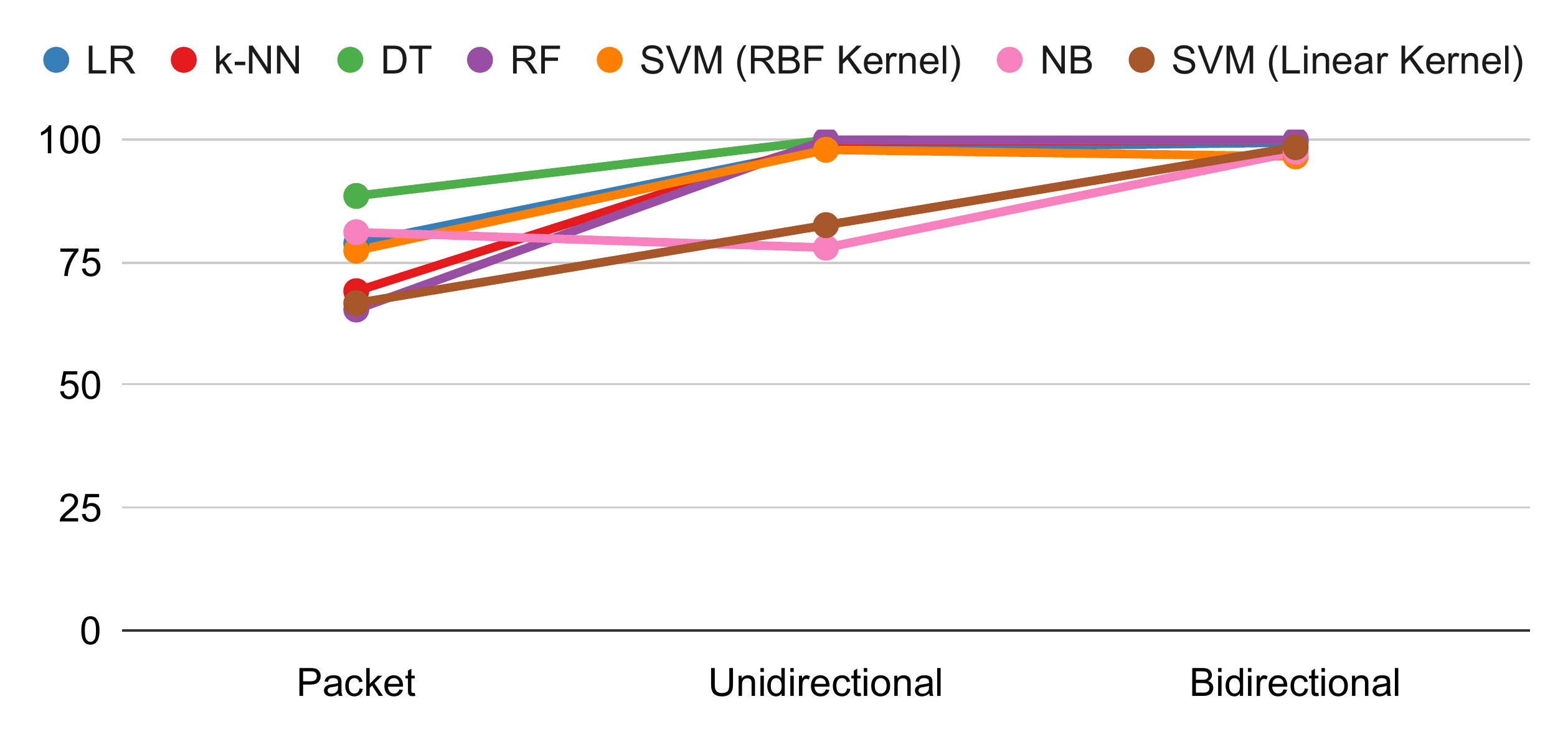}  
  \caption{Recall}
  \label{fig:recall-avg}
\end{subfigure}
\begin{subfigure}{.5\textwidth}
  \centering
  \includegraphics[width=\linewidth]{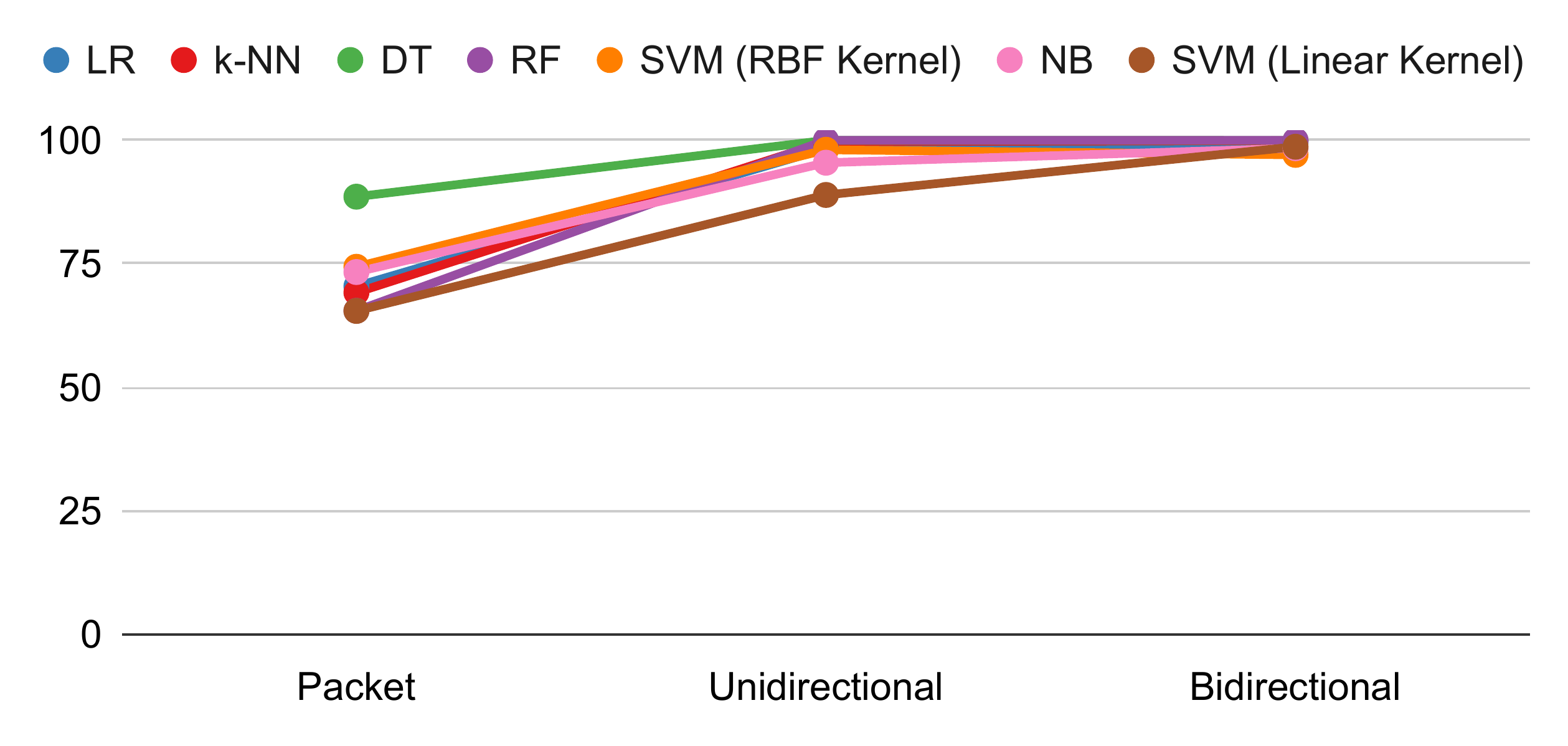}  
  \caption{Precision}
  \label{fig:precision-avg}
\end{subfigure}
\caption{Weighted Average Trends}
\label{fig:avg-trend}
\end{figure}

Furthermore, it is recognised that the Benign and the MQTT-BF attack are the two classes benefiting from flow features. This is reasoned by the fact that in IoT networks normal/benign operations are usually uncomplicated, due to their usage, requirements, and the nature of data of interest. Therefore, generic attacks are quite distinctive. However, MQTT-based attacks have similar characteristics to benign MQTT communication. Since MQTT-based attacks rely on the available MQTT communication commands (i.e., publish, subscribe, etc), it is challenging to discriminate attacks from normal operations where the same commands are used.  
As a result, packet-based features in all the ML techniques were not suitable for benign and MQTT-BF classification. This observation could further be observed in the trends charts for benign class, MQTT\_BF class, and weighted average metrics in Figure~\ref{fig:benign-trend}, Figure~\ref{fig:mqtt-trend} and, Figure~\ref{fig:avg-trend}.

\begin{landscape}
\NetTableWidthTen=\dimexpr
    \linewidth - 20\tabcolsep - 11\arrayrulewidth
\begin{table}[htb]
    \centering
    \begin{tabular}{|C{0.09\NetTableWidthTen} | C{0.09\NetTableWidthTen}| C{0.09\NetTableWidthTen}| C{0.09\NetTableWidthTen}| C{0.09\NetTableWidthTen}| C{0.09\NetTableWidthTen}| C{0.09\NetTableWidthTen}| C{0.09\NetTableWidthTen}| C{0.09\NetTableWidthTen}| C{0.09\NetTableWidthTen}|}
    
\multicolumn{10}{c}{\cellcolor{tabhead} \textbf{LR}}\\ \hline
\multicolumn{1}{c|}{}  & \multicolumn{3}{c|}{\cellcolor{tabhead} \textbf{Recall}} & \multicolumn{3}{c|}{\cellcolor{tabhead} \textbf{Precision}} & \multicolumn{3}{c|}{\cellcolor{tabhead} \textbf{F1-score}} \\ \cline{2-10}
\multicolumn{1}{c|}{}  & \cellcolor{tabhead} \textbf{Packet} & \cellcolor{tabhead} \textbf{Uni} & \cellcolor{tabhead} \textbf{Bi} & \cellcolor{tabhead} \textbf{Packet} & \cellcolor{tabhead} \textbf{Uni} & \cellcolor{tabhead} \textbf{Bi} & \cellcolor{tabhead} \textbf{Packet} & \cellcolor{tabhead} \textbf{Uni} & \cellcolor{tabhead} \textbf{Bi} \\ \hline
\cellcolor{tabhead}Benign & 0\% & 100\% & 99.02\% & 0\% & 93.33\% & 98.95\% & 0\% & 96.55\% & 98.99\% \\ \hline
\cellcolor{tabhead}Scan\_A & 86.45\% & 70.87\% & 97.25\% & 98.39\% & 98.39\% & 97.21\% & 92.03\% & 82.39\% & 97.2\% \\ \hline
\cellcolor{tabhead}Scan\_sU & 98.21\% & 98.03\% & 98.48\% & 99.34\% & 95.76\% & 100\% & 98.77\% & 96.88\% & 99.23\% \\ \hline
\cellcolor{tabhead}Sparta & 100\% & 100\% & 100\% & 98.22\% & 100\% & 100\% & 99.1\% & 100\% & 100\% \\ \hline
\cellcolor{tabhead}MQTT\_BF & 100\% & 99.25\% & 99.58\% & 51.75\% & 99.82\% & 99.41\% & 68.2\% & 99.53\% & 99.5\% \\ \hline
\cellcolor{tabhead}Weighted Average & 78.87\% & 98.23\% & 99.44\% & 70.4\% & 98.32\% & 99.44\% & 72.97\% & 98.14\% & 99.44\% \\ \hline
\multicolumn{10}{c}{}\\ \hline
\multicolumn{10}{c}{\cellcolor{tabhead} \textbf{k-NN}}\\ \hline
\multicolumn{1}{c|}{}  & \multicolumn{3}{c|}{\cellcolor{tabhead} \textbf{Recall}} & \multicolumn{3}{c|}{\cellcolor{tabhead} \textbf{Precision}} & \multicolumn{3}{c|}{\cellcolor{tabhead} \textbf{F1-score}} \\ \cline{2-10}
\multicolumn{1}{c|}{}  & \cellcolor{tabhead} \textbf{Packet} & \cellcolor{tabhead} \textbf{Uni} & \cellcolor{tabhead} \textbf{Bi} & \cellcolor{tabhead} \textbf{Packet} & \cellcolor{tabhead} \textbf{Uni} & \cellcolor{tabhead} \textbf{Bi} & \cellcolor{tabhead} \textbf{Packet} & \cellcolor{tabhead} \textbf{Uni} & \cellcolor{tabhead} \textbf{Bi} \\ \hline
\cellcolor{tabhead}Benign & 17.43\% & 99.69\% & 99.95\% & 17.42\% & 98.85\% & 99.59\% & 17.43\% & 99.27\% & 99.77\% \\ \hline
\cellcolor{tabhead}Scan\_A & 99.99\% & 99.97\% & 100\% & 99.99\% & 99.85\% & 99.9\% & 99.99\% & 99.91\% & 99.95\% \\ \hline
\cellcolor{tabhead}Scan\_sU & 99.99\% & 99.96\% & 100\% & 99.99\% & 99.96\% & 100\% & 99.99\% & 99.96\% & 100\% \\ \hline
\cellcolor{tabhead}Sparta & 100\% & 100\% & 100\% & 100\% & 100\% & 100\% & 100\% & 100\% & 100\% \\ \hline
\cellcolor{tabhead}MQTT\_BF & 25.84\% & 99.3\% & 99.75\% & 25.85\% & 99.82\% & 99.97\% & 25.84\% & 99.56\% & 99.86\% \\ \hline
\cellcolor{tabhead}Weighted Average & 69.13\% & 99.68\% & 99.9\% & 69.13\% & 99.68\% & 99.9\% & 69.13\% & 99.68\% & 99.9\% \\ \hline
\multicolumn{10}{c}{}\\ \hline
\multicolumn{10}{c}{\cellcolor{tabhead} \textbf{DT}}\\ \hline
\multicolumn{1}{c|}{}  & \multicolumn{3}{c|}{\cellcolor{tabhead} \textbf{Recall}} & \multicolumn{3}{c|}{\cellcolor{tabhead} \textbf{Precision}} & \multicolumn{3}{c|}{\cellcolor{tabhead} \textbf{F1-score}} \\ \cline{2-10}
\multicolumn{1}{c|}{}  & \cellcolor{tabhead} \textbf{Packet} & \cellcolor{tabhead} \textbf{Uni} & \cellcolor{tabhead} \textbf{Bi} & \cellcolor{tabhead} \textbf{Packet} & \cellcolor{tabhead} \textbf{Uni} & \cellcolor{tabhead} \textbf{Bi} & \cellcolor{tabhead} \textbf{Packet} & \cellcolor{tabhead} \textbf{Uni} & \cellcolor{tabhead} \textbf{Bi} \\ \hline
\cellcolor{tabhead}Benign & 69.29\% & 99.92\% & 99.88\% & 69.39\% & 99.92\% & 99.91\% & 69.34\% & 99.92\% & 99.9\% \\ \hline
\cellcolor{tabhead}Scan\_A & 100\% & 100\% & 100\% & 99.98\% & 99.95\% & 99.9\% & 99.99\% & 99.97\% & 99.95\% \\ \hline
\cellcolor{tabhead}Scan\_sU & 99.98\% & 99.91\% & 100\% & 100\% & 100\% & 100\% & 99.99\% & 99.96\% & 100\% \\ \hline
\cellcolor{tabhead}Sparta & 100\% & 100\% & 100\% & 100\% & 100\% & 100\% & 100\% & 100\% & 100\% \\ \hline
\cellcolor{tabhead}MQTT\_BF & 72.56\% & 99.95\% & 99.93\% & 72.47\% & 99.95\% & 99.93\% & 72.51\% & 99.95\% & 99.93\% \\ \hline
\cellcolor{tabhead}Weighted Average & 88.55\% & 99.96\% & 99.95\% & 88.55\% & 99.96\% & 99.95\% & 88.54\% & 99.96\% & 99.95\% \\ \hline

    \end{tabular}
    \caption{5-fold cross validation}
    \label{tab:metrics1}
    
\end{table}
\begin{table}[htb]
    \centering
    \begin{tabular}{|C{0.09\NetTableWidthTen} | C{0.09\NetTableWidthTen}| C{0.09\NetTableWidthTen}| C{0.09\NetTableWidthTen}| C{0.09\NetTableWidthTen}| C{0.09\NetTableWidthTen}| C{0.09\NetTableWidthTen}| C{0.09\NetTableWidthTen}| C{0.09\NetTableWidthTen}| C{0.09\NetTableWidthTen}|}
\multicolumn{10}{c}{\cellcolor{tabhead} \textbf{RF}}\\ \hline
\multicolumn{1}{c|}{}  & \multicolumn{3}{c|}{\cellcolor{tabhead} \textbf{Recall}} & \multicolumn{3}{c|}{\cellcolor{tabhead} \textbf{Precision}} & \multicolumn{3}{c|}{\cellcolor{tabhead} \textbf{F1-score}} \\ \cline{2-10}
\multicolumn{1}{c|}{}  & \cellcolor{tabhead} \textbf{Packet} & \cellcolor{tabhead} \textbf{Uni} & \cellcolor{tabhead} \textbf{Bi} & \cellcolor{tabhead} \textbf{Packet} & \cellcolor{tabhead} \textbf{Uni} & \cellcolor{tabhead} \textbf{Bi} & \cellcolor{tabhead} \textbf{Packet} & \cellcolor{tabhead} \textbf{Uni} & \cellcolor{tabhead} \textbf{Bi} \\ \hline
\cellcolor{tabhead}Benign & 9.34\% & 99.96\% & 99.93\% & 8.99\% & 99.94\% & 99.95\% & 9.16\% & 99.95\% & 99.94\% \\ \hline
\cellcolor{tabhead}Scan\_A & 100\% & 100\% & 100\% & 99.98\% & 99.95\% & 99.95\% & 99.99\% & 99.97\% & 99.98\% \\ \hline
\cellcolor{tabhead}Scan\_sU & 99.98\% & 99.91\% & 99.96\% & 99.99\% & 100\% & 100\% & 99.99\% & 99.96\% & 99.98\% \\ \hline
\cellcolor{tabhead}Sparta & 100\% & 100\% & 100\% & 100\% & 100\% & 100\% & 100\% & 100\% & 100\% \\ \hline
\cellcolor{tabhead}MQTT\_BF & 15.15\% & 99.96\% & 99.97\% & 15.69\% & 99.98\% & 99.96\% & 15.42\% & 99.97\% & 99.97\% \\ \hline
\cellcolor{tabhead}Weighted Average & 65.39\% & 99.98\% & 99.97\% & 65.44\% & 99.98\% & 99.97\% & 65.41\% & 99.98\% & 99.97\% \\ \hline
\multicolumn{10}{c}{}\\ \hline
\multicolumn{10}{c}{\cellcolor{tabhead} \textbf{SVM (RBF Kernel)}}\\ \hline
\multicolumn{1}{c|}{}  & \multicolumn{3}{c|}{\cellcolor{tabhead} \textbf{Recall}} & \multicolumn{3}{c|}{\cellcolor{tabhead} \textbf{Precision}} & \multicolumn{3}{c|}{\cellcolor{tabhead} \textbf{F1-score}} \\ \cline{2-10}
\multicolumn{1}{c|}{}  & \cellcolor{tabhead} \textbf{Packet} & \cellcolor{tabhead} \textbf{Uni} & \cellcolor{tabhead} \textbf{Bi} & \cellcolor{tabhead} \textbf{Packet} & \cellcolor{tabhead} \textbf{Uni} & \cellcolor{tabhead} \textbf{Bi} & \cellcolor{tabhead} \textbf{Packet} & \cellcolor{tabhead} \textbf{Uni} & \cellcolor{tabhead} \textbf{Bi} \\ \hline
\cellcolor{tabhead}Benign & 30.23\% & 100\% & 100\% & 28.13\% & 92.67\% & 87.13\% & 28.8\% & 96.19\% & 93.12\% \\ \hline
\cellcolor{tabhead}Scan\_A & 83.8\% & 70.16\% & 42.13\% & 99.99\% & 96.18\% & 99.88\% & 91.18\% & 81.13\% & 59.22\% \\ \hline
\cellcolor{tabhead}Scan\_sU & 92.33\% & 99.96\% & 100\% & 99.74\% & 93.01\% & 94.34\% & 95.89\% & 96.36\% & 97.09\% \\ \hline
\cellcolor{tabhead}Sparta & 100\% & 100\% & 100\% & 91.17\% & 100\% & 100\% & 95.38\% & 100\% & 100\% \\ \hline
\cellcolor{tabhead}MQTT\_BF & 72.42\% & 98.44\% & 98.3\% & 53.56\% & 100\% & 100\% & 59.53\% & 99.22\% & 99.14\% \\ \hline
\cellcolor{tabhead}Weighted Average & 77.4\% & 97.96\% & 96.61\% & 74.35\% & 98.05\% & 97.02\% & 74.89\% & 97.87\% & 96.15\% \\ \hline
\multicolumn{10}{c}{}\\ \hline
\multicolumn{10}{c}{\cellcolor{tabhead} \textbf{NB}}\\ \hline
\multicolumn{1}{c|}{}  & \multicolumn{3}{c|}{\cellcolor{tabhead} \textbf{Recall}} & \multicolumn{3}{c|}{\cellcolor{tabhead} \textbf{Precision}} & \multicolumn{3}{c|}{\cellcolor{tabhead} \textbf{F1-score}} \\ \cline{2-10}
\multicolumn{1}{c|}{}  & \cellcolor{tabhead} \textbf{Packet} & \cellcolor{tabhead} \textbf{Uni} & \cellcolor{tabhead} \textbf{Bi} & \cellcolor{tabhead} \textbf{Packet} & \cellcolor{tabhead} \textbf{Uni} & \cellcolor{tabhead} \textbf{Bi} & \cellcolor{tabhead} \textbf{Packet} & \cellcolor{tabhead} \textbf{Uni} & \cellcolor{tabhead} \textbf{Bi} \\ \hline
\cellcolor{tabhead}Benign & 10.62\% & 1.13\% & 99.96\% & 9.9\% & 97.68\% & 93.56\% & 10.25\% & 2.24\% & 96.65\% \\ \hline
\cellcolor{tabhead}Scan\_A & 100\% & 99.25\% & 66.41\% & 99.23\% & 18.28\% & 100\% & 99.61\% & 30.88\% & 79.81\% \\ \hline
\cellcolor{tabhead}Scan\_sU & 99.52\% & 97.76\% & 100\% & 100\% & 98.79\% & 98.52\% & 99.76\% & 98.27\% & 99.25\% \\ \hline
\cellcolor{tabhead}Sparta & 99.84\% & 100\% & 100\% & 100\% & 100\% & 100\% & 99.92\% & 100\% & 100\% \\ \hline
\cellcolor{tabhead}MQTT\_BF & 90.27\% & 97.78\% & 100\% & 53.15\% & 100\% & 97.05\% & 65.84\% & 98.88\% & 98.5\% \\ \hline
\cellcolor{tabhead}Weighted Average & 81.15\% & 78\% & 97.55\% & 73.29\% & 95.43\% & 98.37\% & 75.99\% & 75.26\% & 97.77\% \\ \hline
    \end{tabular}
    \caption{5-fold cross validation}
    \label{tab:metrics2}
    \end{table}

\begin{table}[htb]
    \centering
    \begin{tabular}{|C{0.09\NetTableWidthTen} | C{0.09\NetTableWidthTen}| C{0.09\NetTableWidthTen}| C{0.09\NetTableWidthTen}| C{0.09\NetTableWidthTen}| C{0.09\NetTableWidthTen}| C{0.09\NetTableWidthTen}| C{0.09\NetTableWidthTen}| C{0.09\NetTableWidthTen}| C{0.09\NetTableWidthTen}|}

\multicolumn{10}{c}{\cellcolor{tabhead} \textbf{SVM (Linear Kernel)}}\\ \hline
\multicolumn{1}{c|}{}  & \multicolumn{3}{c|}{\cellcolor{tabhead} \textbf{Recall}} & \multicolumn{3}{c|}{\cellcolor{tabhead} \textbf{Precision}} & \multicolumn{3}{c|}{\cellcolor{tabhead} \textbf{F1-score}} \\ \cline{2-10}
\multicolumn{1}{c|}{}  & \cellcolor{tabhead} \textbf{Packet} & \cellcolor{tabhead} \textbf{Uni} & \cellcolor{tabhead} \textbf{Bi} & \cellcolor{tabhead} \textbf{Packet} & \cellcolor{tabhead} \textbf{Uni} & \cellcolor{tabhead} \textbf{Bi} & \cellcolor{tabhead} \textbf{Packet} & \cellcolor{tabhead} \textbf{Uni} & \cellcolor{tabhead} \textbf{Bi} \\ \hline
\cellcolor{tabhead}Benign & 57.34\% & 99.84\% & 99.26\% & 27.8\% & 58.95\% & 97.45\% & 37.38\% & 73.82\% & 98.32\% \\ \hline
\cellcolor{tabhead}Scan\_A & 83.28\% & 68.23\% & 84.1\% & 70.42\% & 70.35\% & 93.44\% & 69.7\% & 67.5\% & 87.01\% \\ \hline
\cellcolor{tabhead}Scan\_sU & 78.13\% & 60.31\% & 97.76\% & 75.8\% & 70.71\% & 93.77\% & 76.92\% & 61.91\% & 95.27\% \\ \hline
\cellcolor{tabhead}Sparta & 87.64\% & 60.37\% & 99.99\% & 97.62\% & 99.94\% & 100\% & 89.89\% & 74.61\% & 99.99\% \\ \hline
\cellcolor{tabhead}MQTT\_BF & 24.89\% & 97.79\% & 98.71\% & 43.3\% & 99.89\% & 99.55\% & 20.84\% & 98.83\% & 99.13\% \\ \hline
\cellcolor{tabhead}Weighted Average & 66.69\% & 82.6\% & 98.5\% & 65.42\% & 88.9\% & 98.66\% & 60.4\% & 82.42\% & 98.46\% \\ \hline

    \end{tabular}
    \caption{5-fold cross validation}
    \label{tab:metrics3}
    \end{table}
    
\end{landscape}

\section{Conclusion and Future Work}
\label{sec:conclusion}

This work aims at exploring the different challenges and requirements for building IDS for IoT models, using an MQTT network as a case study. This paper evaluates six different ML techniques as attack classifiers. 
A simulated MQTT network was used for data collection to simulate a real-life setup.  
Using the dataset raw pcap files, three features levels were extracted; packet, unidirectional, and bidirectional features.
Each feature level is used independently in the experiments.
The experiments highlighted that generic networking attacks are easily discriminated from normal operation due to their distinguished behaviour and patterns compared to the IoT setup. However, MQTT-based attacks are more complicated and can easily mimic benign operation.

The experimental results further demonstrated that the flow-based features are better suited to discriminate between benign and MQTT-based attacks due to their similar characteristics. The weighted average recall rose from $\sim75.31\%$ for packet-based features to $\sim93.77\%$ and $\sim98.85\%$ for unidirectional and bidirectional flow features, respectively. While the weighted average precision rose from $\sim72.37\%$ for packet-based features to $\sim97.19\%$ and $\sim99.04\%$ for unidirectional and bidirectional flow features.
Therefore, the experiments emphasised on the special challenges faced by IoT IDS, based on their custom communication patterns. The challenges were  demonstrated through the difficulties to differentiate MQTT-based attacks from normal operations.

\bibliographystyle{splncs04}
\bibliography{bibliography}

\end{document}